\documentclass[twoside,slac_one]{revtex4}
\usepackage{graphicx}
\usepackage{fancyhdr}
\usepackage{amsmath} 
\usepackage{bm}
\usepackage{amsxtra}
\usepackage{amssymb}
\usepackage{amsthm}
\usepackage{latexsym}
\usepackage{lscape}

\def\T0{\hbox{$t_0$}}
\def\RT{\hbox{$r-t$}}

\def\pt{\hbox{$p_T$}}

\begin{document}

\title{Calibration and Performance of the ATLAS Muon Spectrometer}

%

\author{E. Diehl on behalf of ATLAS}
\affiliation{Department of Physics, University of Michigan, Ann Arbor,
  MI, USA}

\begin{abstract}
The ATLAS muon spectrometer is designed to measure muon momenta with a
resolution of 4\% @ 100 GeV/c rising to 10\% @ 1 TeV/c track
momentum. The spectrometer consists of precision tracking and trigger
chambers embedded in a 2T magnetic field generated by three large
air-core superconducting toroids. The precision detectors provide 50
$\mu$m tracking resolution to a pseudo-rapidity of 2.7. The system
also includes an optical monitoring system which measures detector
positions with 40 $\mu$m precision. This paper reports on the
calibration and performance of the ATLAS muon spectrometer.
\end{abstract}

\maketitle

\thispagestyle{fancy}


\section{Introduction}

The ATLAS experiment\cite{atlas_paper} is one of two general purpose
collider detectors for the Large Hadron Collider (LHC) at CERN.  The
ATLAS detector consists of an inner detector employing silicon pixel,
strip, and transition radiation tracking detectors, all in a
solenoidal magnetic field of 2 Tesla; electromagnetic and hadronic
calorimeters using liquid argon and scintillator tile detectors; and a
muon spectrometer.  The muon spectrometer consists of a large air-core
barrel and endcap toroid magnets with a $B \cdot dl$ between 2-6 $T
\cdot m$, and four types of trigger and precision tracking detectors,
described below.  The muon spectrometer is designed to measure the
transverse momentum ($\pt$) of muons with $\pt > 3$ GeV with a
resolution of 4\% up to $\pt$ of 100 GeV and increasing to 10\% @ 1
TeV.

The ATLAS muon spectrometer consists of Monitored Drift Tubes (MDTs)
for precision tracking in the spectrometer bending plane, Resistive
Plate Chambers (RPCs) and Thin Gap Chambers (TGCs) for triggering in
barrel and endcap, respectively, and Cathode Strip Chambers (CSCs) for
precision measurements in the high-rate endcap inner layer where MDTs
would have occupancy problems.

The magnet system consists of 3 sets of air-core toroids, each with 8
coils, 1 for the barrel, and 1 for each endcap.  The barrel toroids
coils are each 25m $\times$ 7m and the endcap coils are 9m $\times$
4m.  The magnetic field provides an approximately 1T field at the
center of each coils, but is non-uniform, especially in the
barrel-endcap transition region.  For track reconstruction, the field
is mapped using a computer model of the field which is normalized to
measurements from 1850 Hall sensors mounted on spectrometer chambers.

Alignment measurements of the spectrometer are also critical for
momentum determination and are accomplished with an optical alignment
system of 12k sensors.  Measurements from these sensors allow a
3-dimensional reconstruction of chamber positions accurate to better
than 50 $\mu$m.  In addition, the optical alignment system is
complemented by alignment done with tracks.

\begin{figure}[ht]
\includegraphics[width=80mm]{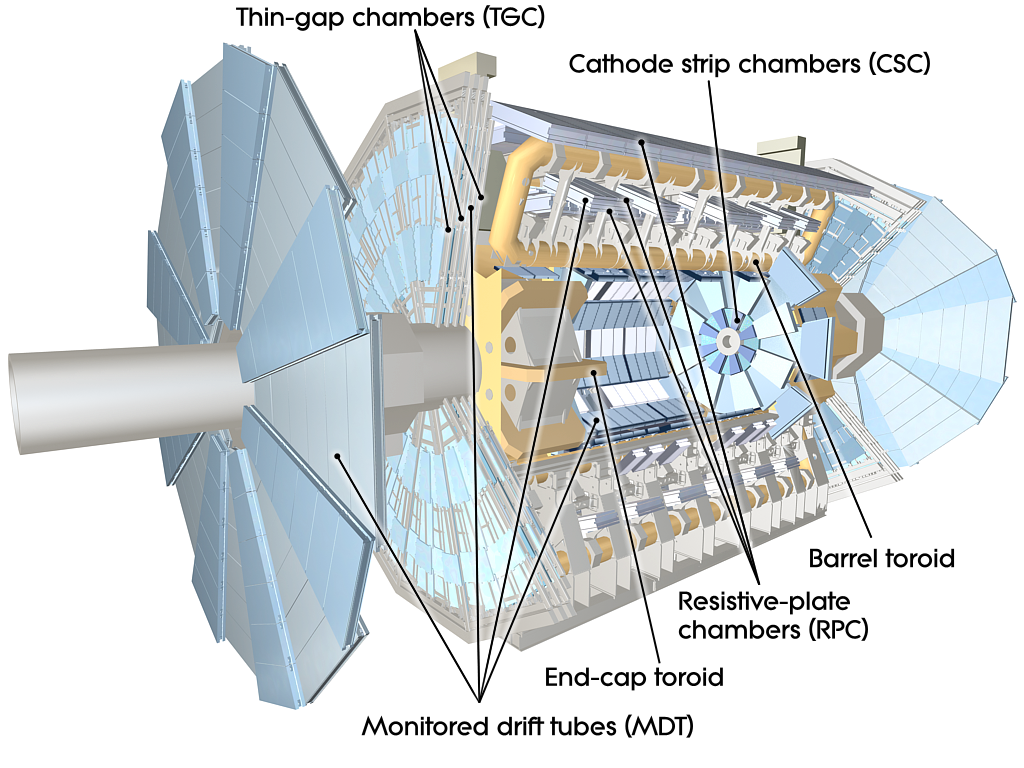}
\includegraphics[width=80mm]{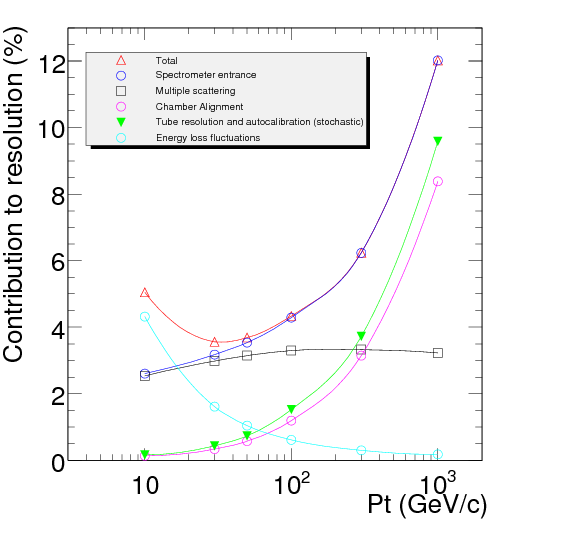}
\caption{Left: Layout of the muon spectrometer; Right: Expected
  resolution of the ATLAS muon spectrometer.}
 \label{fig:layout}
\end{figure}

Table \ref{table:muon_spectrometer} gives a summary of the muon
spectrometer detector components and Fig. \ref{fig:layout} shows
the layout.

\begin{table}[h]
\begin{center}
\caption{ATLAS Muon Spectrometer.}
\begin{tabular}{|l|c|c|c|c|}
\hline \textbf{Type} & \textbf{Purpose} & \textbf{location} &
\textbf{$\eta$ coverage} & \textbf{Channels}  \\
\hline 
MDT & Tracking & barrel+endcap      & $0.0 < \eta < 2.7$ & 354k \\
CSC & Tracking & endcap layer 1     & $2.0 < \eta < 2.7$ & 30.7k \\
RPC & Trigger  & barrel             & $0.0 < \eta < 1.0$ & 373k \\
TGC & Trigger  & endcap             & $1.0 < \eta < 2.4$ & 318k \\
\hline
\end{tabular}
\label{table:muon_spectrometer}
\end{center}
\end{table}

The spectrometer is designed so that muons cross three layers of MDT
chambers for the sagitta measurement.  The track coordinate in the
bending plane of the spectrometer is measured by the precision
chambers with a resolution of 40 $\mu$m.  In comparison, the sagitta
of a 1 TeV muon will be about 500 $\mu$m.  The trigger chambers are
placed on opposite sides of the middle MDT layer.  The trigger
chambers provide a trigger based on muon momentum in addition to
identifying the bunch crossing time of the muon.  The also provide the
second coordinate measurement (non-bending plane) accurate to 5-10 cm.
 
Figure \ref{fig:layout} shows the expected resolution of the muon
spectrometer.  For $\pt < 100$ GeV/c multiple scattering is the
dominant contributor to the resolution.  Above 100 GeV/c calibration
and alignment of the spectrometer become the most significant factors
in momentum resolution.  ATLAS muon reconstruction is done using
momentum measurements from both the inner detector spectrometer and
the muon spectrometer.  The two spectrometers nicely complement each
other as inner detector measurements are better below 100 GeV/c above
which the muon spectrometer resolution is superior.  High $\pt$
measurements with the muon spectrometer require very accurate MDT and
alignment calibrations which will become particularly important in a
few years when LHC reaches 7 TeV energy per beam and higher
luminosities.

This paper shows a mix of results from 2010 and 2011 using the most
up-to-date plots whenever possible.  Not all plots are available for
2011 so in some cases 2010 plots are used.

\section{Monitored Drift Tube Calibrations}

There are three calibrations required for the MDTs: timing offsets
($\T0$); time-space ($\RT$) functions; and drift tube resolution
functions\cite{calib_paper}.  In order to obtain high quality
calibrations for the MDTs a special high statistics calibration data
stream is extracted second-level trigger processors and sent for
processing at three calibration centers at Michigan, Rome, and Munich.
This calibration stream provides 10-100X the rate of single muon
tracks compared to regular ATLAS data.  With this stream it is
possible to do daily calibrations of the monitored drift tubes as well
as detailed data quality monitoring.

\subsection{Timing Offset Calibrations}

\begin{figure}[ht]
\includegraphics[width=80mm]{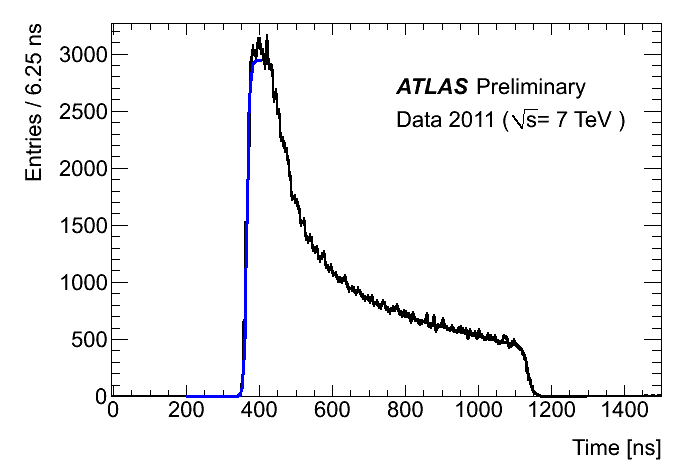}
\includegraphics[width=80mm]{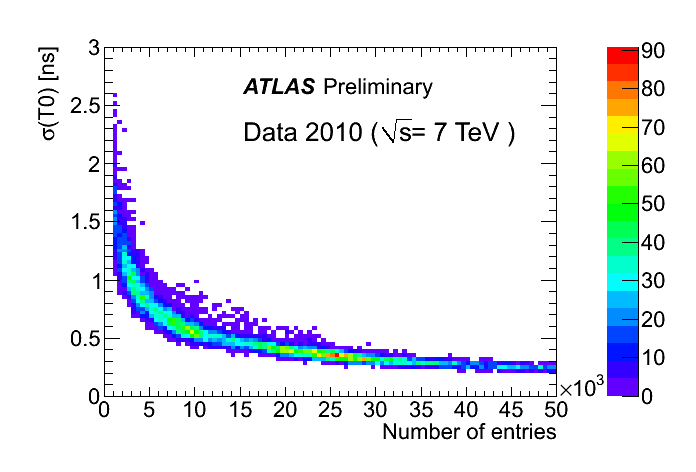}
\caption{Left: An MDT drift time spectrum with $\T0$ fit shown in
  blue; Right: The statistical error on $\T0$ fits as a function of
  number hits in the time spectrum. } 
\label{fig:drifttime}
\end{figure}

A timing offset represents the minimum measured drift time. i.e.
the time of a muon passing at the wire of the drift tube.  This time
is not zero due to cables and other delays in the data acquisition
system.  Figure \ref{fig:drifttime} shows a typical time spectrum
from an MDT.  The $\T0$ fit is shown in blue, and the $\T0$ is
defined as the half-way point of the rising edge.  The falling edge
represents hits at the tube wall. Figure \ref{fig:drifttime} shows
statistical error on $\T0$ fits as a function of number hits in the
time spectrum.  We require at least 10000 hits for the fits yielding a
typical error of 0.5 ns.  The average drift speed is 20 $\mu$m/ns
so this error corresponds to a 10 $\mu$m error due to the $\T0$
measurement.

\begin{figure}[ht]
\includegraphics[width=89mm]{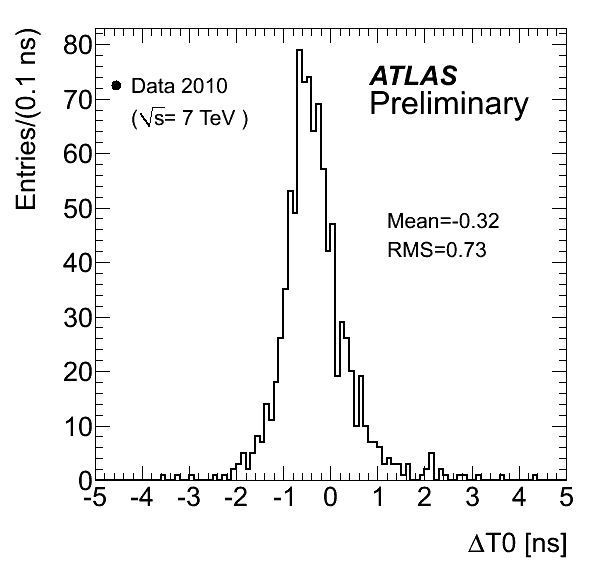}
\includegraphics[width=71mm]{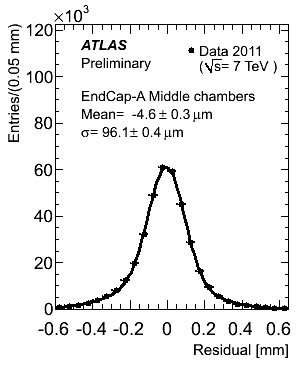}
\caption{Left: The difference between $\T0$s for all MDT chambers over a
  2-month period in 2010. Right: Tracking residuals from the middle layer of
  MDT chambers. } 
\label{fig:t0diff}
\end{figure}

Figure \ref{fig:t0diff} shows the distribution of the change in $\T0$
for all MDT chambers over a 2 month period from 2010.  There is small
global drift of a fraction of a nanosecond, but the overall width is
close to the typical statistical error of 0.5 ns.  Hence, the T0s are
quite stable.
  
\subsection{$\RT$ function calibration}

\begin{figure}[ht]
\includegraphics[width=65mm]{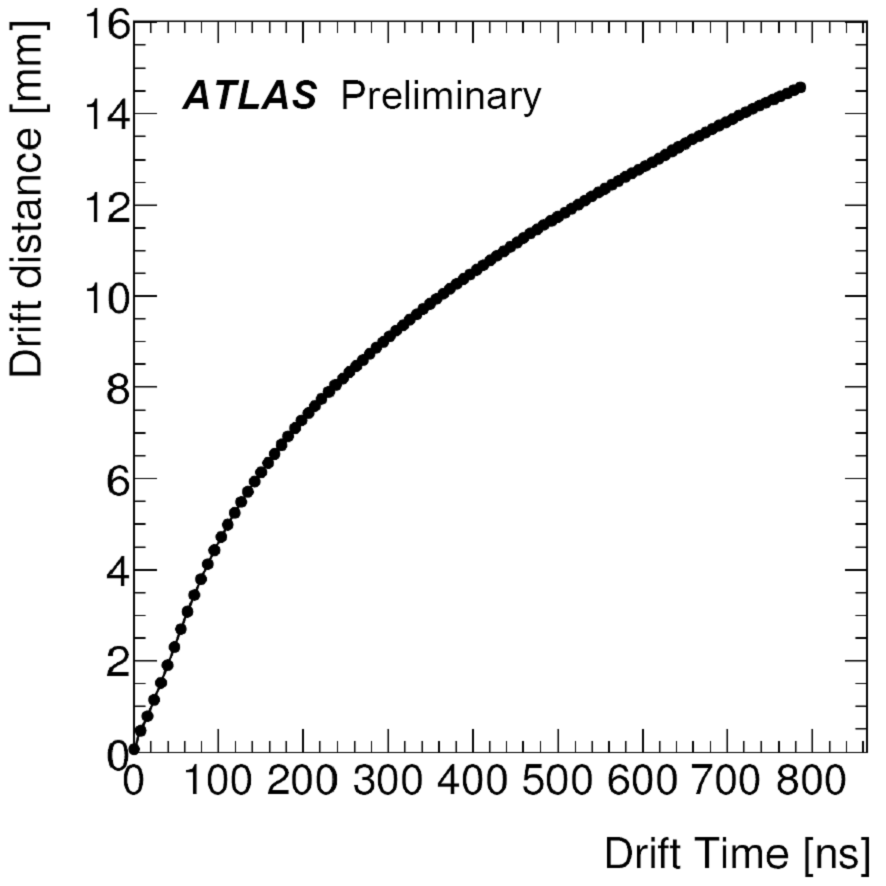}
\includegraphics[width=99mm]{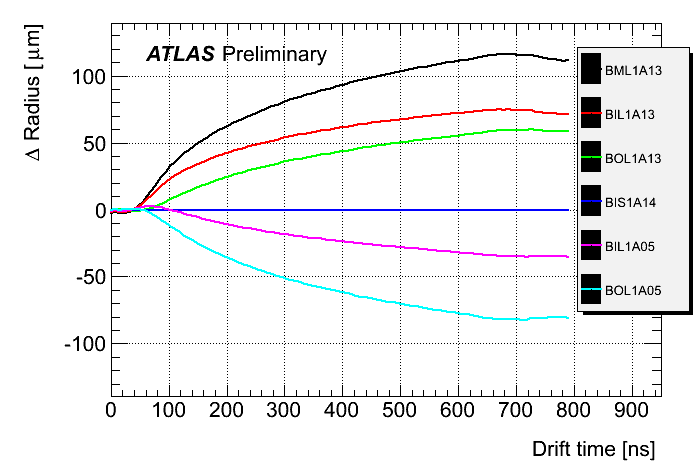}
\caption{Left: A typical $\RT$ function; Right: differences between
  $\RT$ functions for several barrel chambers.} 
\label{fig:rtfun}
\end{figure}

The other main calibration is the time-to-space or $\RT$ function.
This function gives the drift radius (impact parameter) of the hit
based on the drift time.  An example of an $\RT$ function is shown in
Fig. \ref{fig:rtfun}.  The function is non-linear since ATLAS uses a
non-linear drift gas, $\rm ArCO_2$ due to its better aging
characteristics in high-radiation environments.

The $\RT$ function is determined by an iterative procedure looping
over tubes hits and minimizing the tracking residuals of track
segments reconstructed with hits within a single MDT chamber which
have either 6 or 8 layers of MDT tubes.  Tracking residuals are the
differences between the drift radii from the drift time and the radius
from the track fit.  Typically 10000 tracks segments are used in the
calibration of a single chamber.  Figure \ref{fig:rtfun} shows the
difference between $\RT$ functions for several chambers in the barrel.
The differences in $\RT$ functions are due primarily to the
temperature gradient within the ATLAS cavern (about $20 ^{\circ}C$
from top to bottom), as well as due differences in magnetic field
within chambers.

\begin{figure}[ht]
\includegraphics[width=90mm]{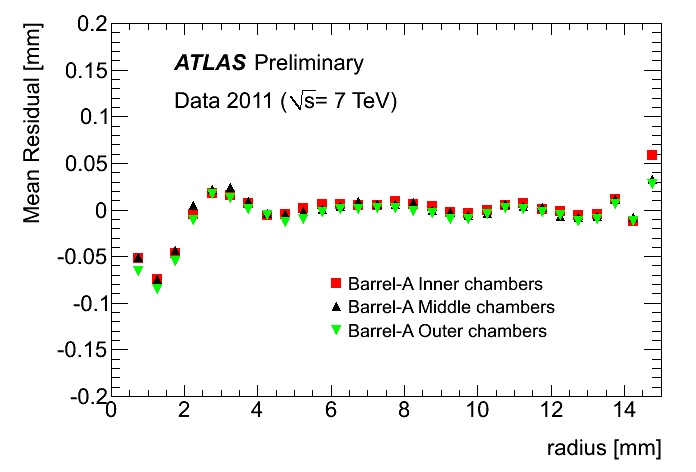}
\includegraphics[width=70mm]{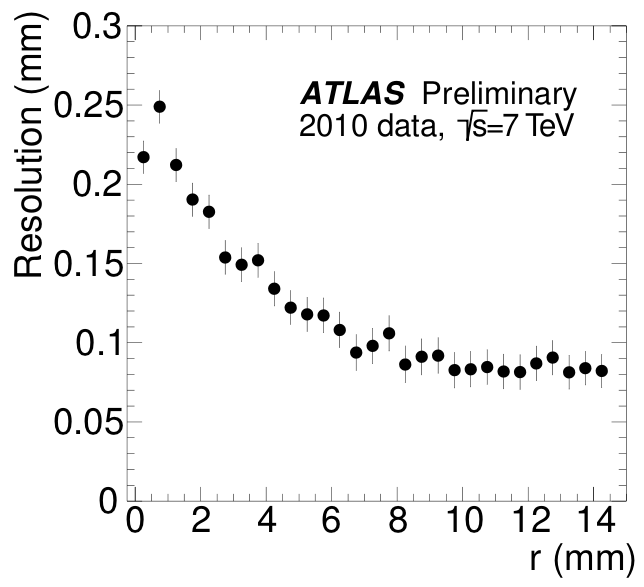}
\caption{Left: MDT tube residuals mean as a function of tube radius;
  Right: MDT hit resolution as a function of tube radius.} 
\label{fig:resvr}
\end{figure}

The precision of the $\RT$ function is shown by Fig. \ref{fig:resvr}
which shows the mean of the tracking residuals as a function of the
MDT tube radius.  Except for the region close to the wire, the mean
residuals are within 20 $\mu$m.  Tracking residuals from the middle
layer of endcap chambers are shown in Fig. \ref{fig:t0diff}.  The
residual width of 96 $\mu$m is typical for all chambers in ATLAS.

Figure \ref{fig:resvr} shows the single tube resolution as a
function of drift tube radius.  The resolution is determined from the
tracking residuals width with the fit errors subtracted.  We see that
the resolution is close to the 80 $\mu$m for large radii.  Near the
wire the resolution degrades to faster drift speed and fewer drift
electrons.  This plot was made with 2010 data.  We expect improvements
in the future by applying some addition timing corrections such as a
hit-level magnetic field correction and by using tube-level $\T0$s.

\section{Muon Spectrometer Alignment}

The alignment system system is designed to track chamber positions
with a 40 $\mu$m precision.  The monitoring is done with optical
sensors which is cross-checked by doing alignment with straight tracks
from magnet-off runs.  The barrel and endcap have separate alignment
systems.

\begin{figure}[ht]
\includegraphics[width=80mm]{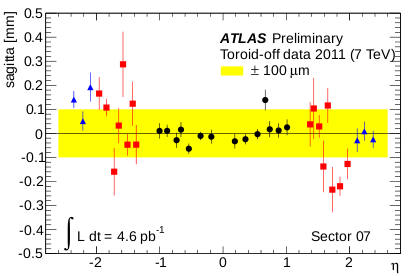}
\caption{Alignment resolution of $\phi$ sector 7 of the muon spectrometer.} 
\label{fig:alignment}
\end{figure}

Figure \ref{fig:alignment} shows measurements from the mean value of
the "false" sagitta measured with straight tracks from magnet-off
runs.  Straight tracks should have a sagitta of zero and hence this
sagitta measurement gives the precision of the alignment system.  The
plot shows the sagitta as a function of $\eta$ with the black points
for the barrel and the red and blue corresponding with to the endcap.
The barrel achieves a resolution of 50 $\mu$m, close to the design
goal, whereas the endcap gives a resolution of around 110 $\mu$m
indicting that further improvements are necessary.

\section{Trigger Performance}

\begin{figure}[ht]
\includegraphics[width=80mm]{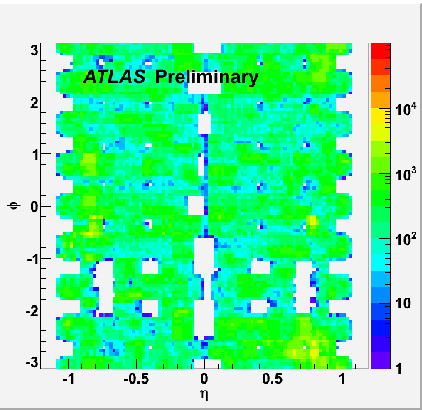}
\caption{Trigger occupancy for the barrel as a function of $\phi$ and $\eta$.}
\label{fig:triggerocc}
\end{figure}

\begin{figure}[ht]
\includegraphics[width=80mm]{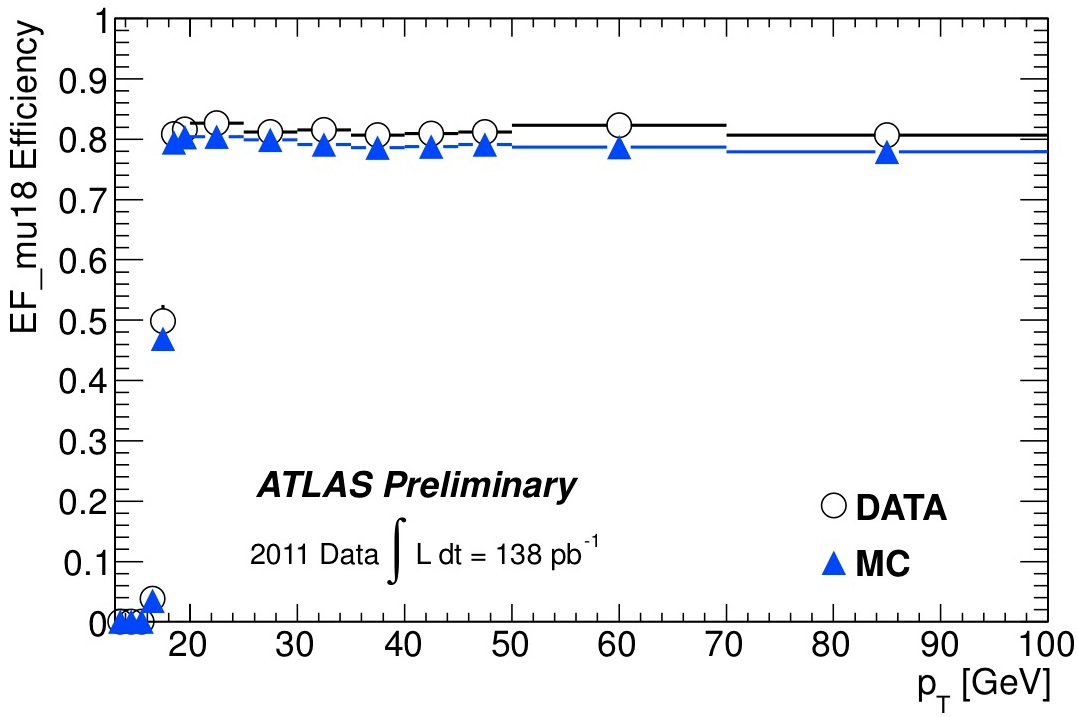}
\includegraphics[width=80mm]{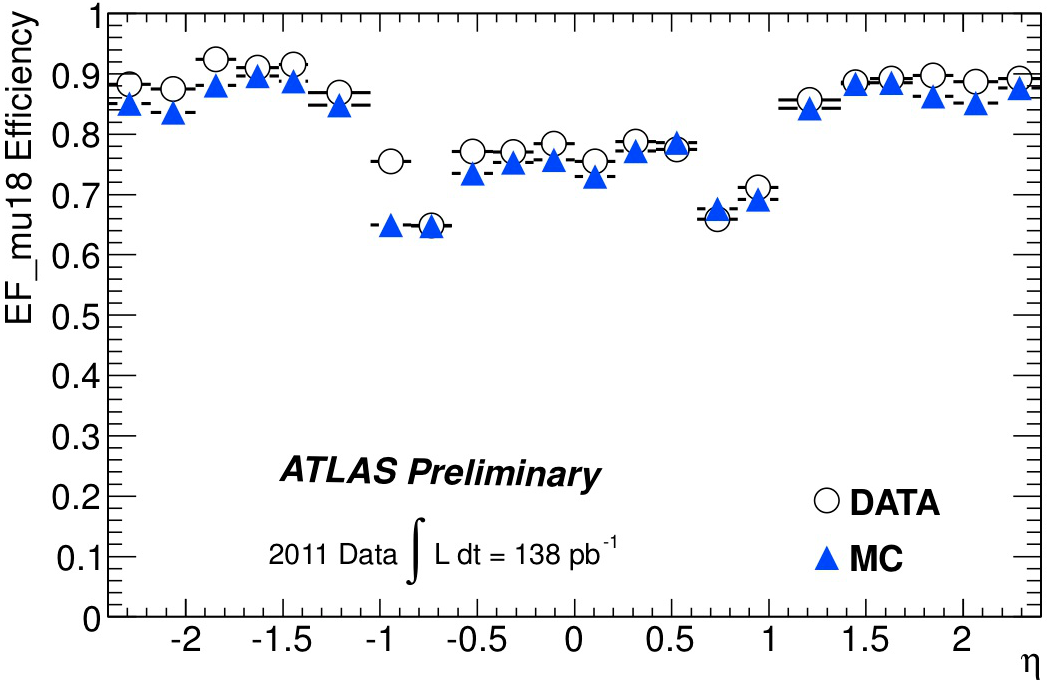}
\caption{Left: Trigger efficiency as a function of $\pt$; Right:
  Trigger efficiency as a function $\eta$.}
\label{fig:triggereff}
\end{figure}

Figure \ref{fig:triggerocc} shows the an occupancy plot for the barrel
from collision data.  As can be seen the coverage is quite uniform
except for dead regions due to the support feet of the ATLAS detector.
The geometric acceptance is about 80\%.  Figure \ref{fig:triggereff}
shows the trigger efficiency as a function of $\pt$ and $\eta$,
respectively.  The triggers are very efficient within geometric
coverage of the trigger.  The data show a slightly higher efficiency
than monte carlo due to some analysis improvements in the data
analysis which have not yet been introduced to the monte carlo code.

\section{Momentum Resolution}

\begin{figure}[ht]
\includegraphics[width=80mm]{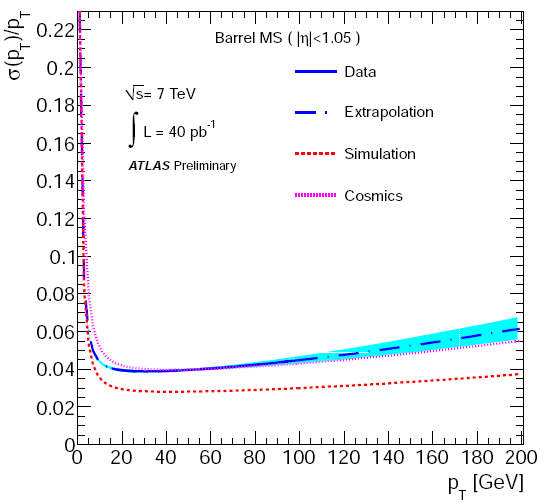}
\caption{Momentum resolution of the ATLAS muon spectrometer barrel region.} 
\label{fig:momresb}
\end{figure}

Figure \ref{fig:momresb} shows the momentum resolution as a function
of muon $\pt$ in the barrel region of ATLAS.  The momentum resolution
is derived from the width of the reconstructed Z mass as well as by
comparing single muons reconstructed by both the inner detector and
muon spectrometer.  Monte carlo and inner detector measurements are
used to derive the contributions of momentum resolution from energy
loss in calorimeters, multiple scattering, and the intrinsic
resolution of the spectrometer.  The measured resolution is somewhat
worse than the simulation.  This result is from the preliminary
calibrations for 2010 data, so we expect improvement from more refined
calibrations in the future.

\section{Conclusions}

The ATLAS muon spectrometer is working well with a high trigger
efficiency and tracking resolution near design specifications.
Calibrations of the drift tubes are done daily using a high statistics
data stream from the level-2 trigger processors.  Alignment is working
well in barrel, but needs some improvement in the endcap.  Momentum
resolution is near design specifications.  We expect improvements from
better calibrations and statistics from 2011 data.

\bigskip 

\end{document}